\documentclass{PoS}

\title{Overview of latest results from PHENIX}

\ShortTitle{PHENIX Overview}

\author{\speaker{Takao Sakaguchi for the PHENIX collaboration}\\
        Brookhaven National Laboratory\\
        E-mail: \email{takao@bnl.gov}}

\abstract{An overview of the latest results on the hard probes from the PHENIX experiment at RHIC is given. The results on the measurements of high $p_{T}$ hadrons, hadron-hadron correlations, open heavy flavor and quarkonia, and direct photons from large (Au+Au) to small collision systems ($p$+Al and $p/d/^3$He+Au) provided a deeper insight on the medium created in the large systems and the possible onset of QGP-nization in transition from small to large systems. }

\FullConference{International Conference on Hard and Electromagnetic Probes of High-Energy Nuclear Collisions\\
		30 September - 5 October 2018\\
		Aix-Les-Bains, Savoie, France}

\begin{document}

\section{Introduction}
The Relativistic Heavy Ion Collider (RHIC) at Brookhaven National Laboratory (BNL)
has been operated almost for two decades, during which variety of nucleus have
been collided at various energies, in addition to  the golden colliding mode of
Au+Au collisions at $\sqrt{s_{NN}}$~=~200\,GeV. In the last five years of data
taking and analysis, PHENIX focused on not only large systems like Au+Au
collisions, but also small systems like $p$+Al and $p/d/^3$He+Au collisions. We will show
the latest results from these systems and discuss what we learned.

\section{Results from large system}
\begin{figure}[htbp]
\centering
\includegraphics[width=0.6\textwidth]{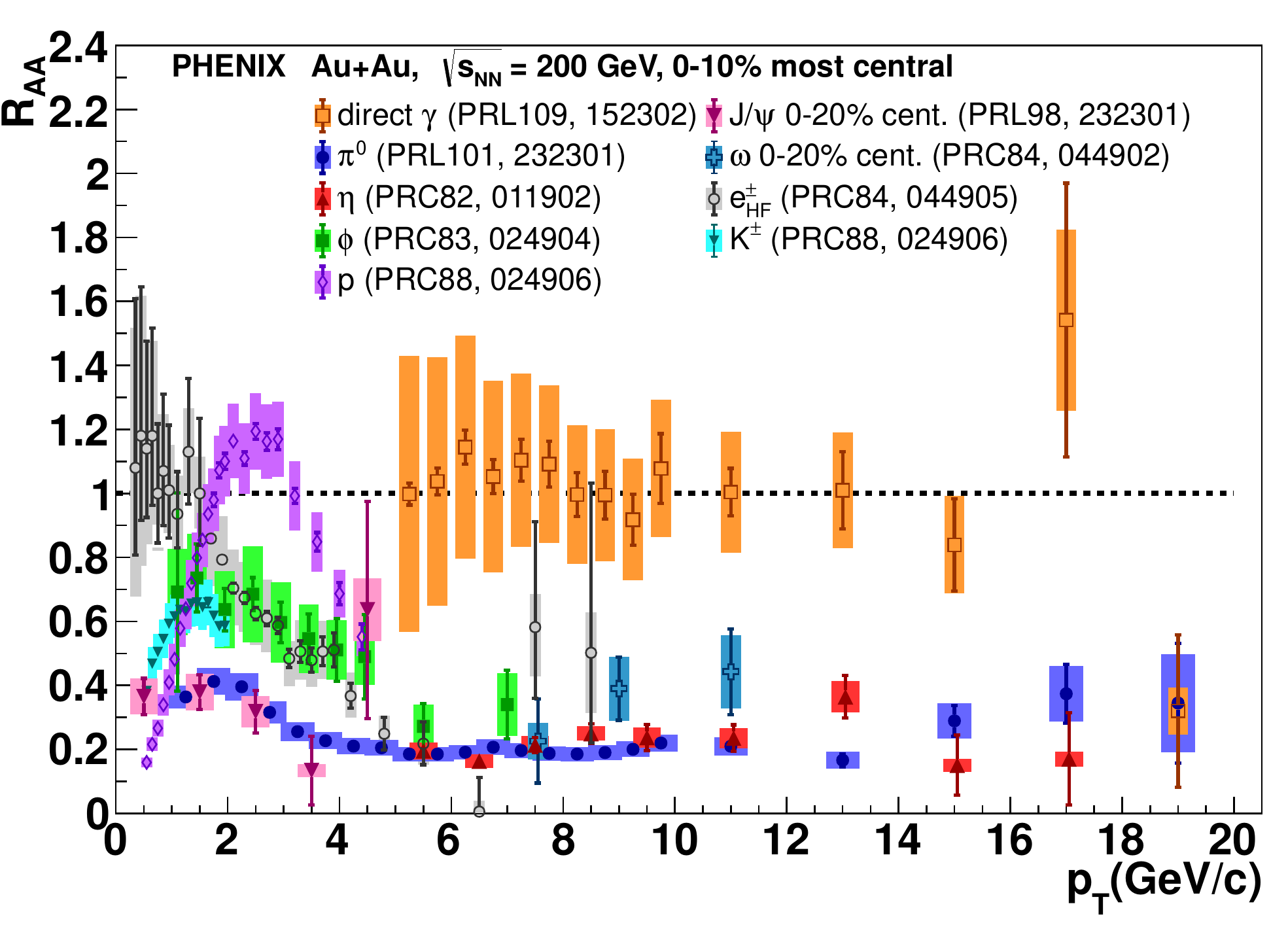}
\caption{Latest compilation of $R_{\rm AA}$ for various particles in 0-10\,\% Au+Au collisions at $\sqrt{s_{NN}}$~=~200\,GeV.}
\label{fig0}
\end{figure}

\begin{figure}[htbp]
\centering
\begin{minipage}{0.44\textwidth}
\includegraphics[width=1.0\textwidth]{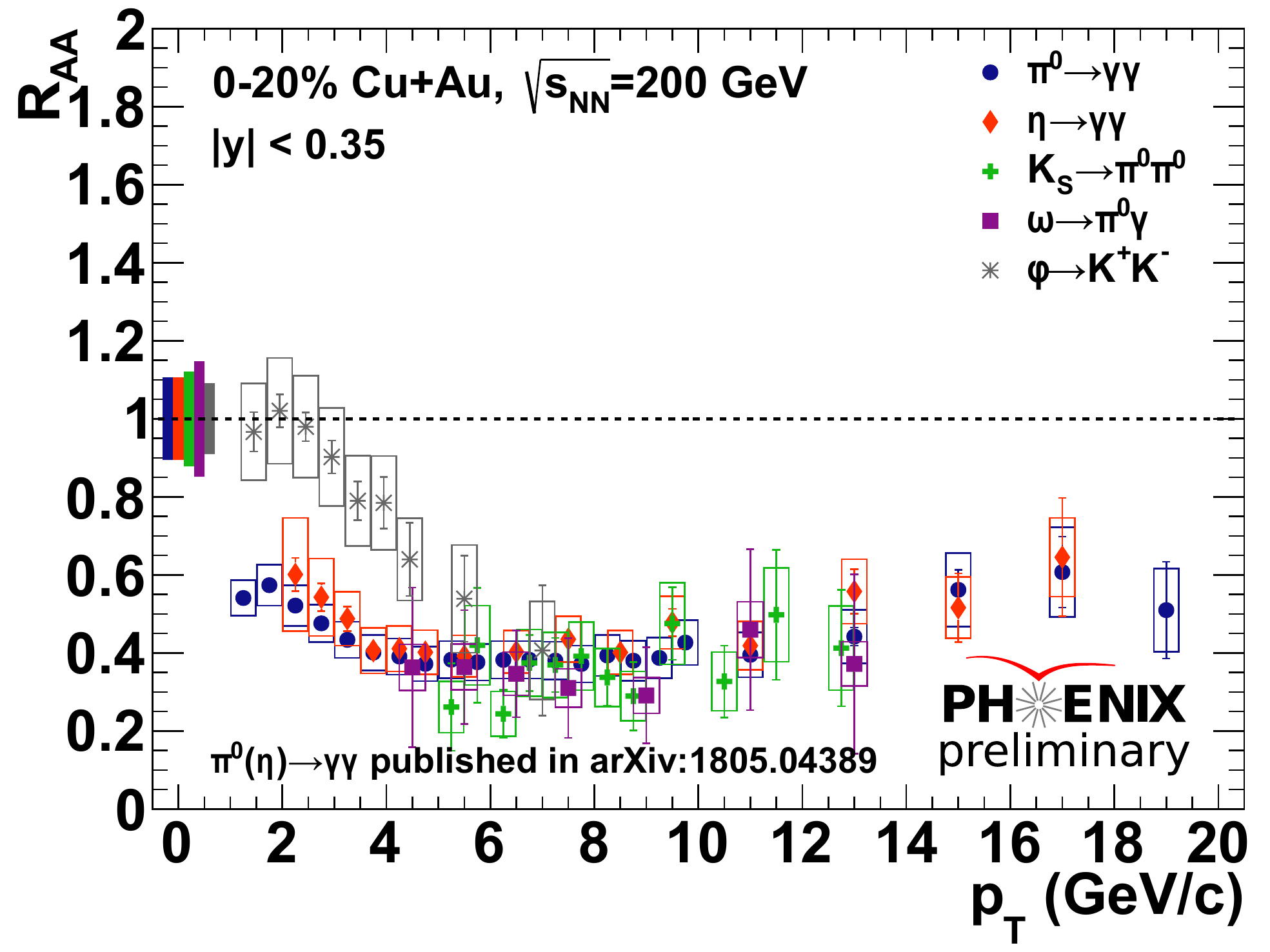}
\end{minipage}
\hspace{1mm}
\begin{minipage}{0.54\textwidth}
\includegraphics[width=1.0\textwidth]{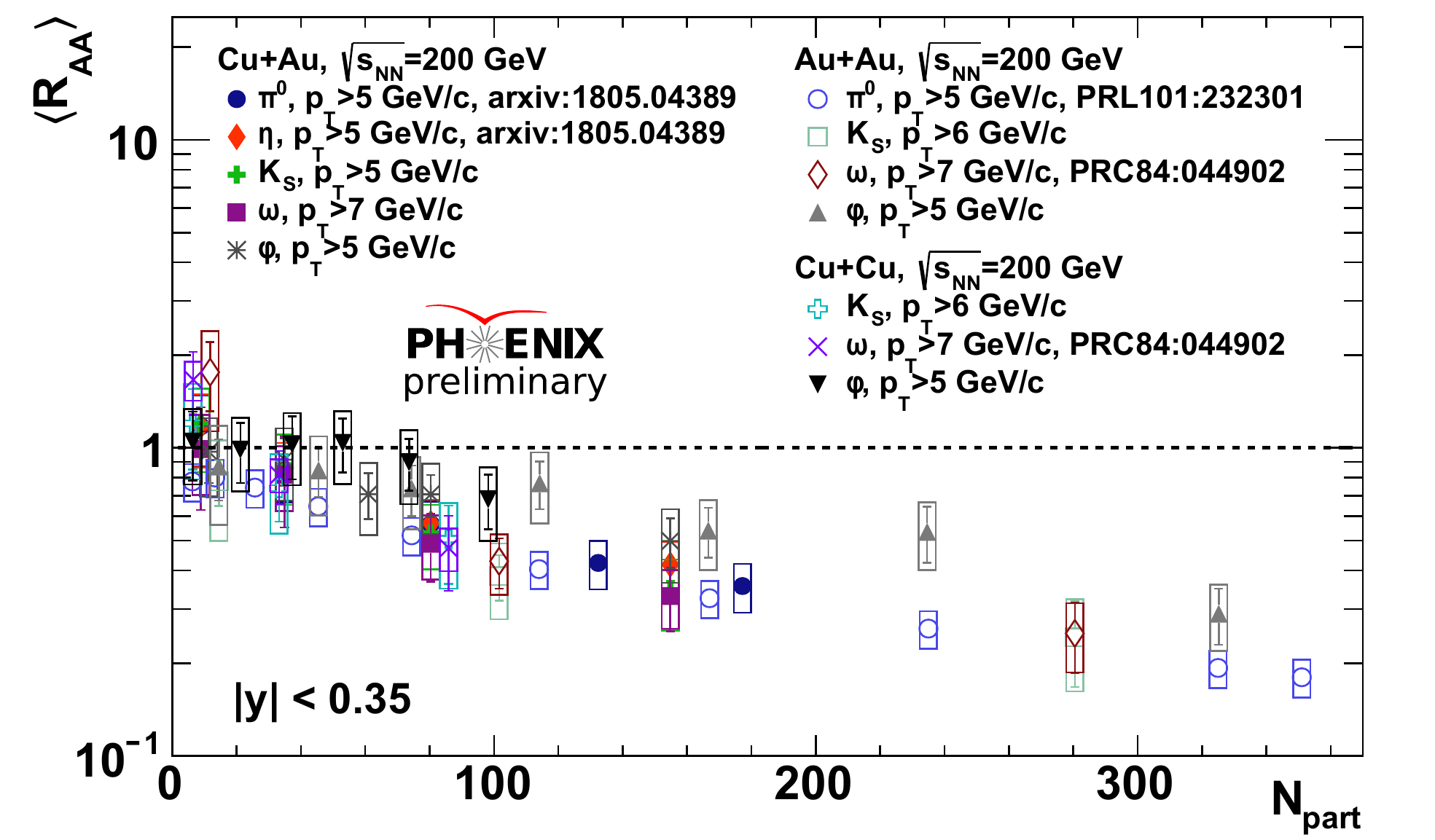}
\end{minipage}
\caption{(a, left) $R_{\rm AA}$ for hadrons in 0-20\,\% Cu+Au collisions at
$\sqrt{s_{NN}}$~=~200\,GeV. (b, right) Integrated $R_{\rm AA}$ for
Au+Au, Cu+Au and Cu+Cu collisions for $p_{T}$ above 5--7\,GeV/$c$, as
a function of $N_{\rm part}$.}
\label{fig1}
\end{figure}

From the beginning of the experiment, the measurement of the high $p_{T}$
hadrons has been of our primary focus.
Figure~\ref{fig0} shows the latest compilation of the nuclear modification
factors ($R_{\rm AA}$) for various particles emitted in 0-10\,\% Au+Au
collisions at $\sqrt{s_{NN}}$~=~200\,GeV.
It is clearly
seen that the yields of light mesons are equally suppressed over $p_{T}$,
except for $\phi$ at low $p_{T}$, while the direct
photons are consistent with the expectation from the primordial production.
With the Cu+Au collisions performed in the RHIC Year-2012 run, we have
extend the compilation to an asymmetric system, as shown in Fig.~\ref{fig1}(a).
The yields of light hadrons ($\pi^0$, $\eta$, $\omega$) are again equally
suppressed, while a strange hadron ($\phi$) is off the trend
in $p_{T}<5$\,GeV/$c$, which is consistent with the Au+Au result.
When looking at the integrated $R_{\rm AA}$ for $p_{T}>5-7$\,GeV/$c$ as a
function of $N_{\rm part}$ as shown in Fig.~\ref{fig1}(b), the $R_{\rm AA}$ values follow a common trend. This is consistent with the fact that the
$\phi$ is also equally suppressed when going to higher $p_{T}$~\cite{ref1}.

\begin{figure}[htbp]
\centering
\begin{minipage}{0.52\textwidth}
\includegraphics[width=1.0\textwidth]{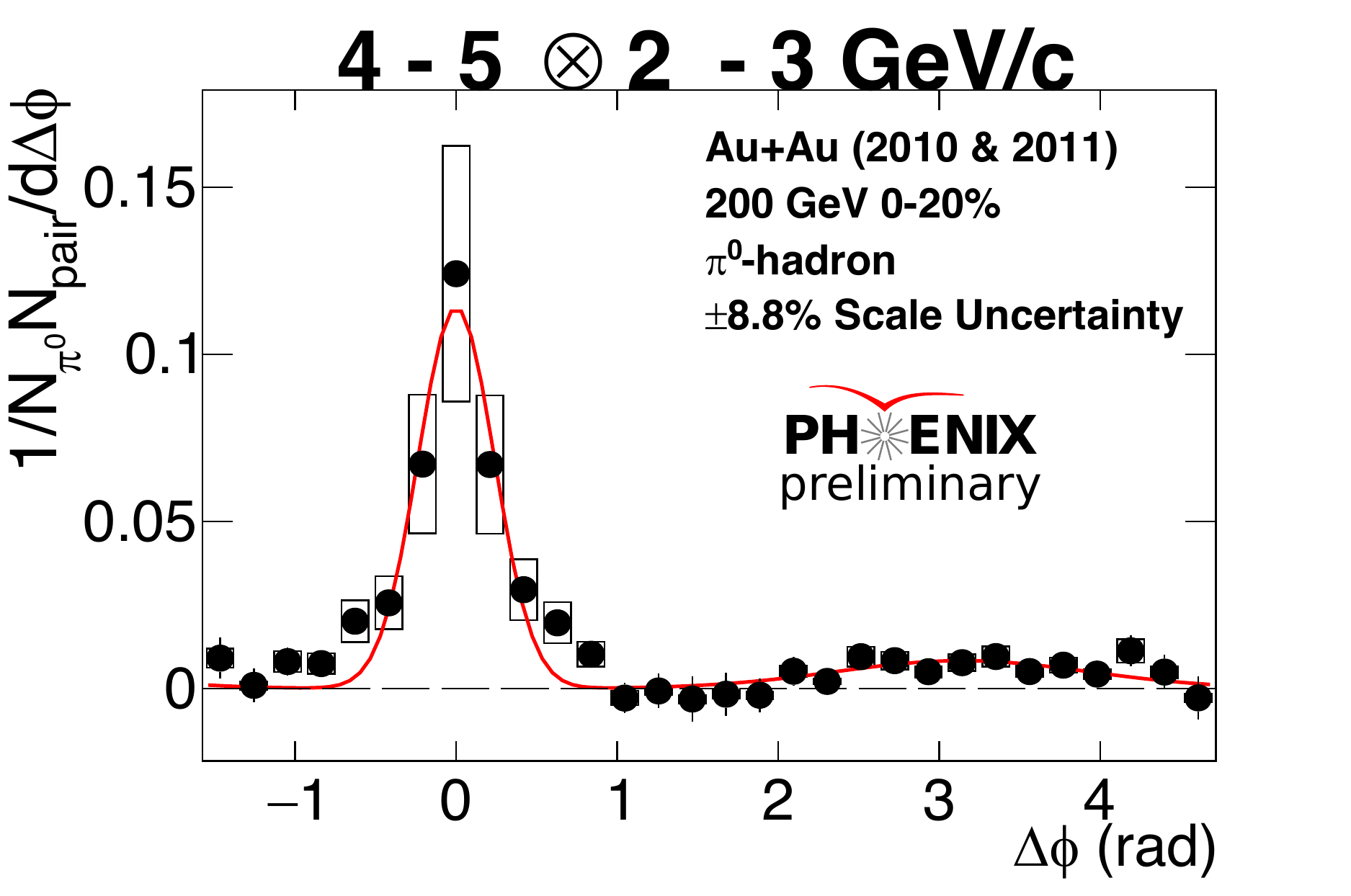}
\end{minipage}
\hspace{2mm}
\begin{minipage}{0.43\textwidth}
\includegraphics[width=1.0\textwidth]{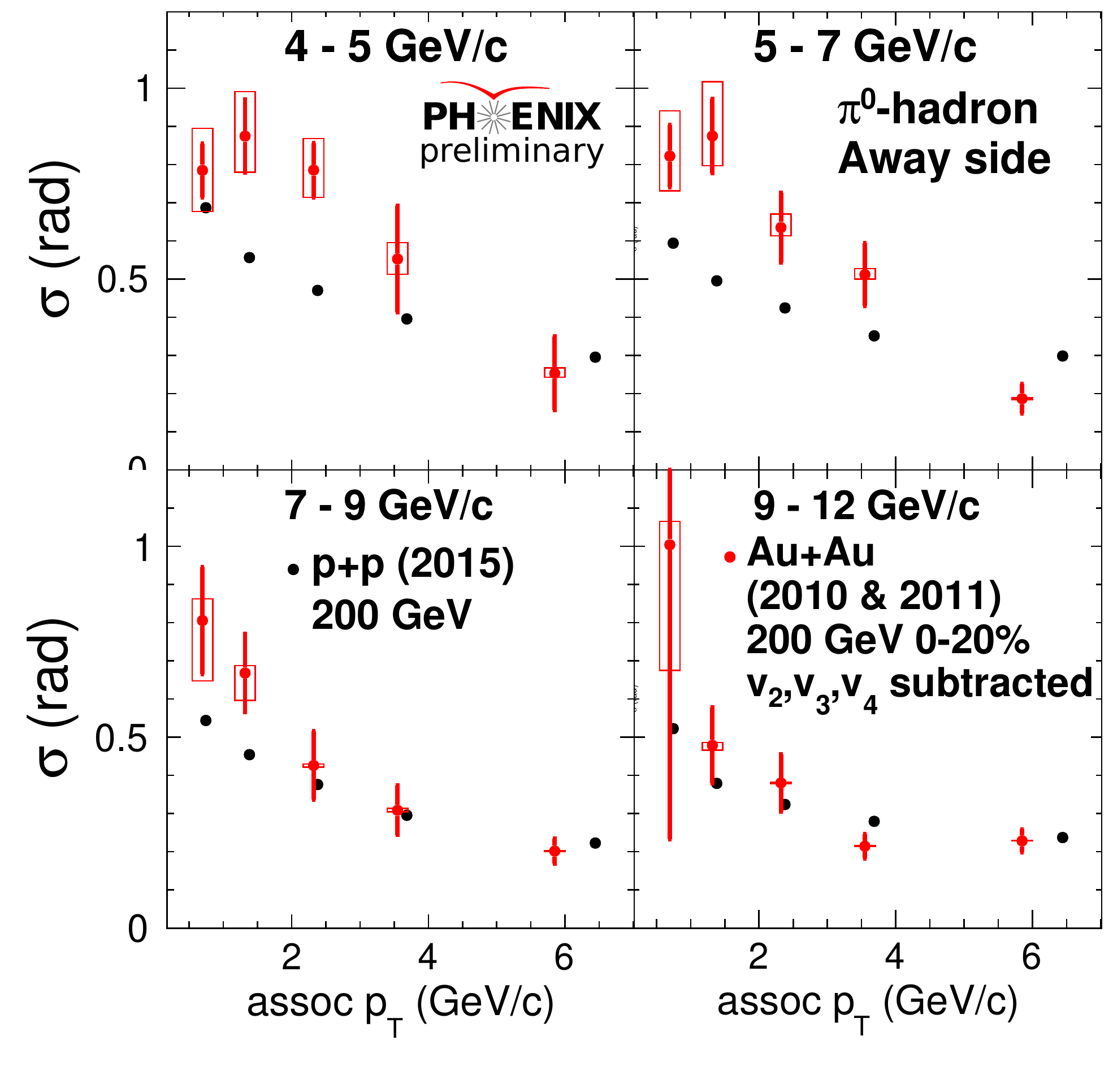}
\end{minipage}
\caption{(a, left) Jet function from $\pi^0$-$h$ correlations in 0-20\,\% Au+Au
collisions at $\sqrt{s_{NN}}$~=~200\,GeV. (b, right) Away-side peak widths of the
jet functions, as a function of trigger $\pi^0$ $p_{T}$ and associated
hadron $p_{T}$.}
\label{fig2}
\end{figure}

The hadron-hadron correlation gives us additional insight of the medium
compared to single hadrons. PHENIX has measured the $\pi^0$-hadron correlations
in Au+Au collisions at $\sqrt{s_{NN}}$~=~200\,GeV in the past, and obtained the
width of the near-side and away-side peak of jet functions~\cite{ref2}.
At that time, the particle flow was explored up to second order ($v_2$),
therefore the background flow subtraction was performed only up to the second
order as well. With the high
statistics RHIC Year-2010 and 2011 run data and taking $v_n$ ($n=2,3,4$) flow
components into account for background estimate, the jet functions are significantly
improved and smooth in $\Delta\phi$ as shown in Fig.~\ref{fig2}(a).
The widths of the away-side peaks are shown in Fig.~\ref{fig2}(b). Comparing to
the previous result, both the statistical and systematic uncertainties are
much improved, which results in a firmer conclusion that the widths are larger
for Au+Au collisions compared to that for $p$+$p$ at low $p_{T}$, and
they converge as going to higher $p_{T}$~\cite{ref3}. The result can be
compared to the $\gamma$-hadron correlation result whose trigger particles
don't interact with medium~\cite{ref4}.

\begin{figure}[htbp]
%\vspace{-5mm}
\centering
\begin{minipage}{0.51\textwidth}
\centering
\includegraphics[width=0.9\textwidth]{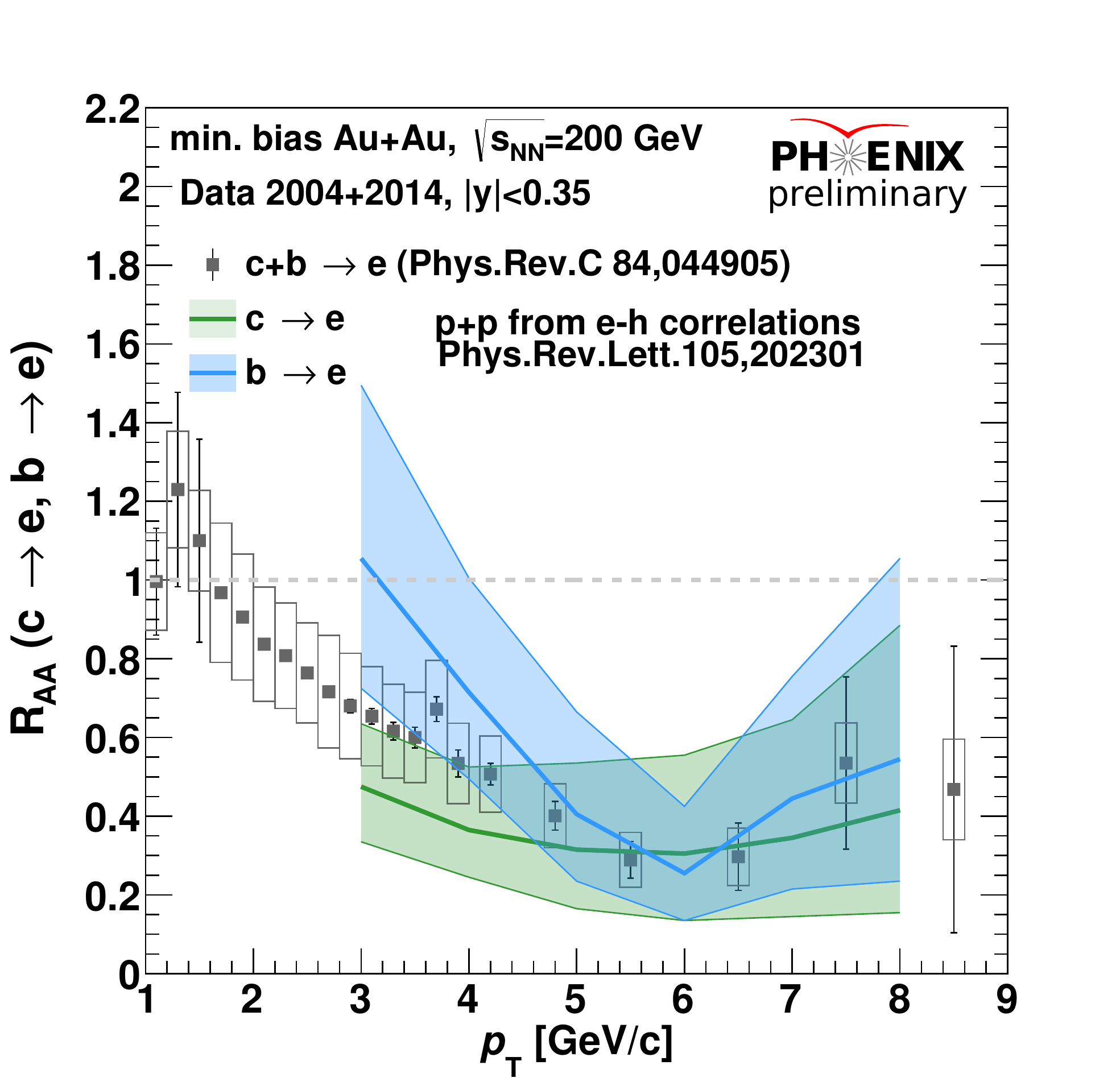}
\end{minipage}
\begin{minipage}{0.46\textwidth}
\centering
\includegraphics[width=0.9\textwidth]{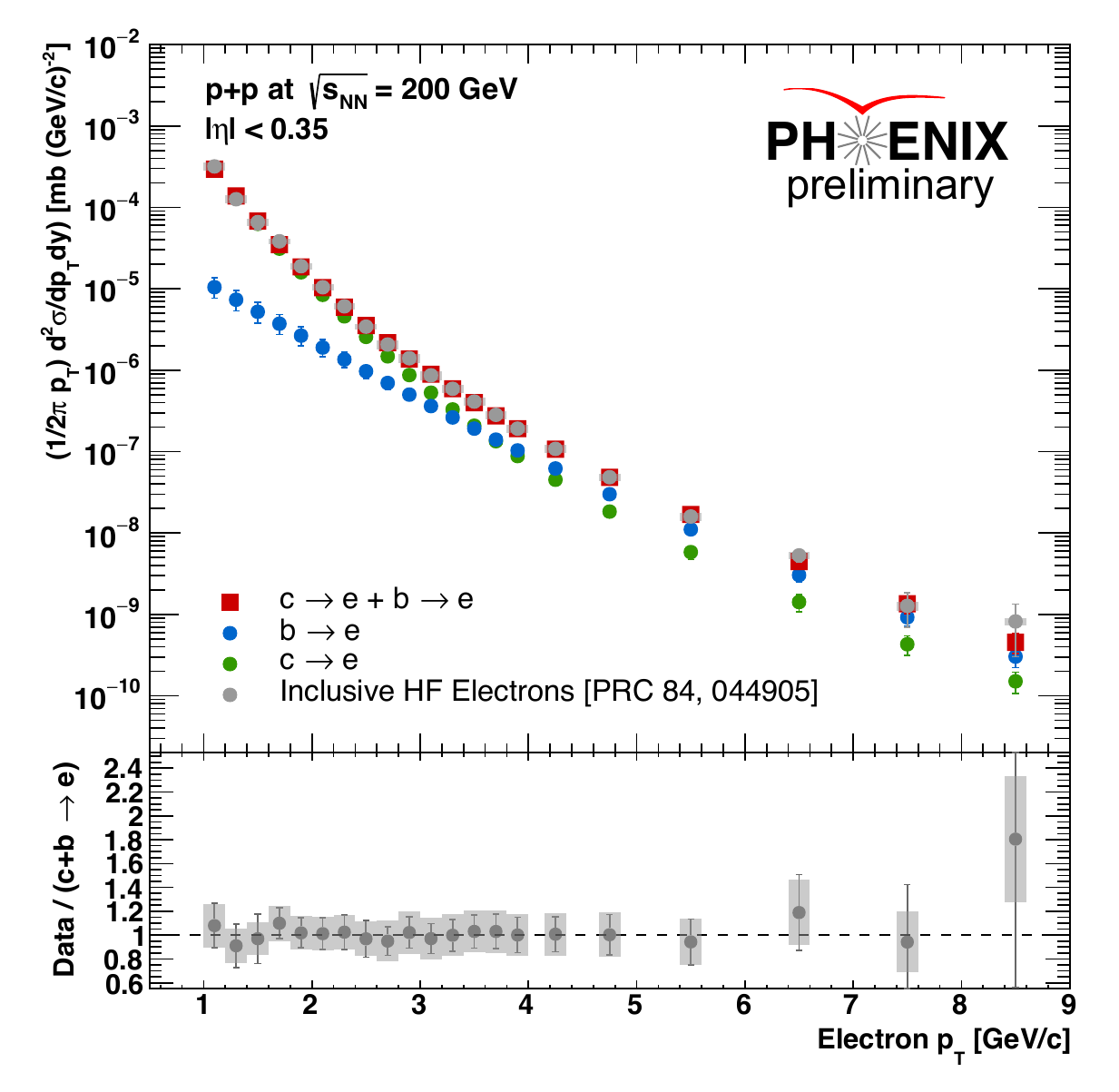}
\end{minipage}
\caption{(a, left) $R_{\rm AA}$ for inclusive electrons from charm and
bottom quarks, and electrons from charm and bottom quarks separately,
measured in minimum bias Au+Au collisions at $\sqrt{s_{NN}}$~=~200\,GeV.
(b, right) New $p$+$p$ baseline of the electrons from charm and bottom quarks.}
\label{fig3}
\end{figure}

\begin{figure}[htbp]
\centering
\begin{minipage}{0.48\textwidth}
\centering
\includegraphics[width=1.0\textwidth]{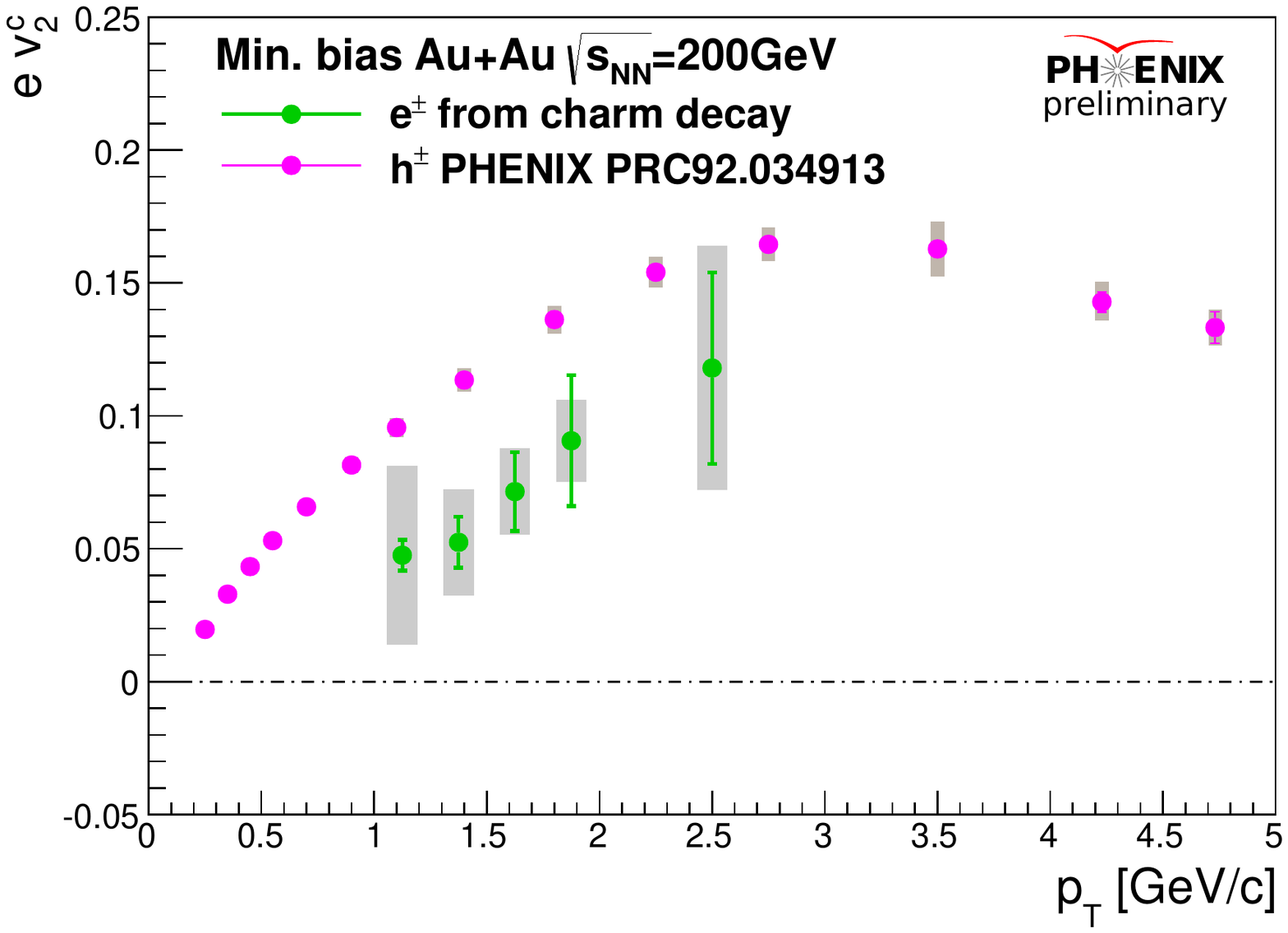}
\end{minipage}
\begin{minipage}{0.48\textwidth}
\centering
\includegraphics[width=1.0\textwidth]{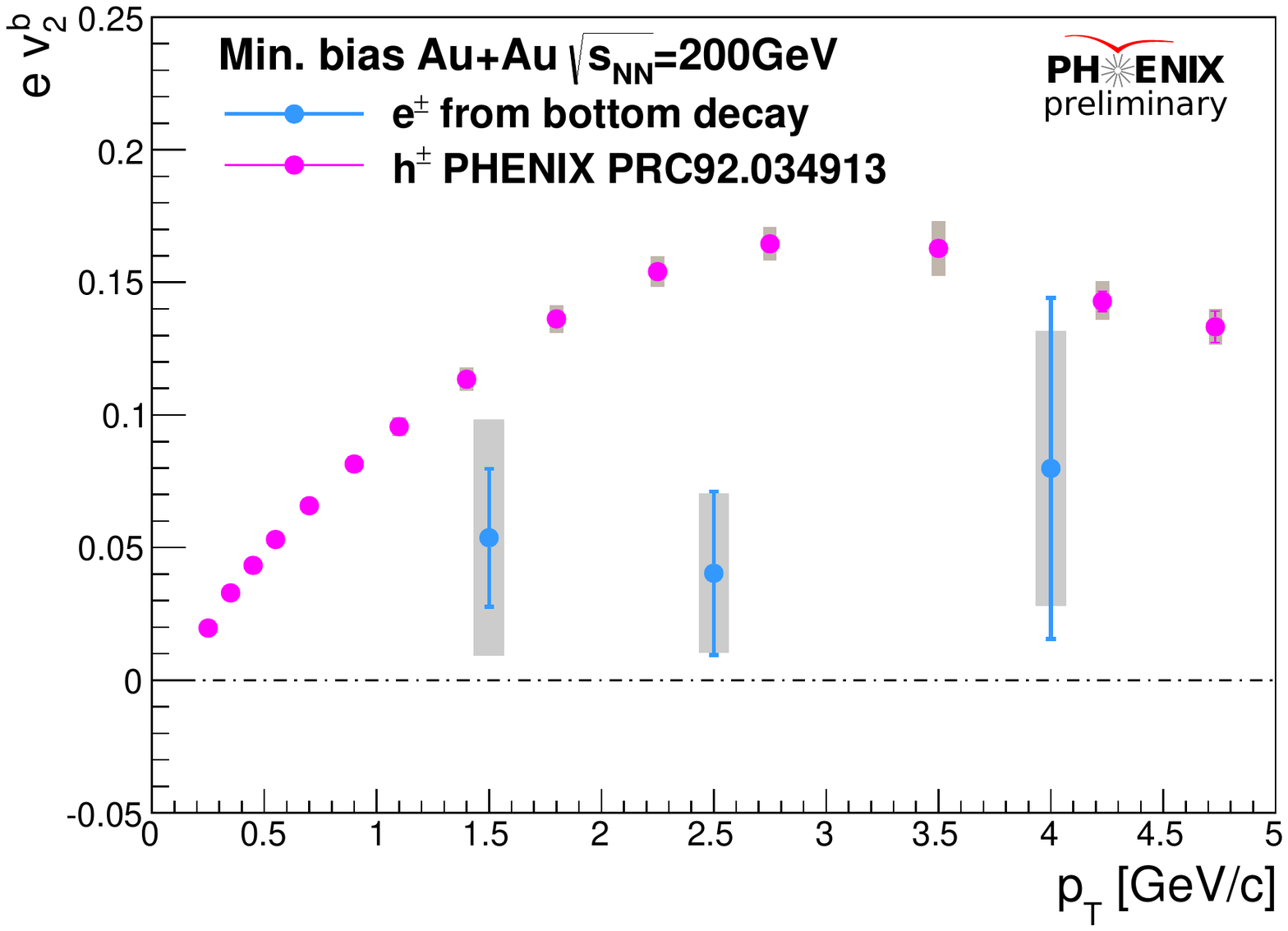}
\end{minipage}
\caption{$v_2$ of the inclusive heavy flavor electrons compared with
(a, left) unfolded charm electrons and (b, right) bottom electrons.}
\label{fig4}
\end{figure}

A systematic study of the energy loss as a function of quark mass gives
another handle on the energy loss mechanism. PHENIX has measured
electrons and muons from heavy flavor quark decay (charm and bottom) and
unfolded to each component. Figure~\ref{fig3}(a) shows the $R_{\rm AA}$
for the inclusive heavy flavor electrons, together with the electrons
from charm and bottom separately in minimum bias Au+Au collisions at
$\sqrt{s_{NN}}$~=~200\,GeV.
A hint of the mass ordering in the suppression is seen; electrons from
bottom quarks tend to be less suppressed compared to those from charm
quarks. The large errors, however, prevented us from making a definitive
conclusion. A dominant source of uncertainty in this
measurement is the
fact that the $p$+$p$ reference was made up from the $e$-$h$ correlation
result by the STAR experiment. With the RHIC Year-2015
data which has the VTX detector, PHENIX succeeded
to measure the electrons from charm and bottom quarks separately
in $p$+$p$ collisions,
as shown in Fig.~\ref{fig3}(b)~\cite{ref5}. A forthcoming
$R_{\rm AA}$ measurement will use this new $p$+$p$ baseline. 
If the heavy quarks lose their energies significantly, they may eventually
stop in the medium and follow the expansion of the bulk system, in which
case these quarks will flow. PHENIX has also successfully measured the
flow of electrons from charm and bottom quarks
separately, as shown in Fig.~\ref{fig4}.
The electrons from bottom quarks seem to flow less than those from
charm quarks. Together with $R_{\rm AA}$, the result implies that
less energy loss of a heavy quark leads to less probability of the
quark being stopped and merged into the bulk system~\cite{ref6}.

\section{Transition from large to small systems}
\begin{figure}[htbp]
\centering
\includegraphics[width=0.8\textwidth]{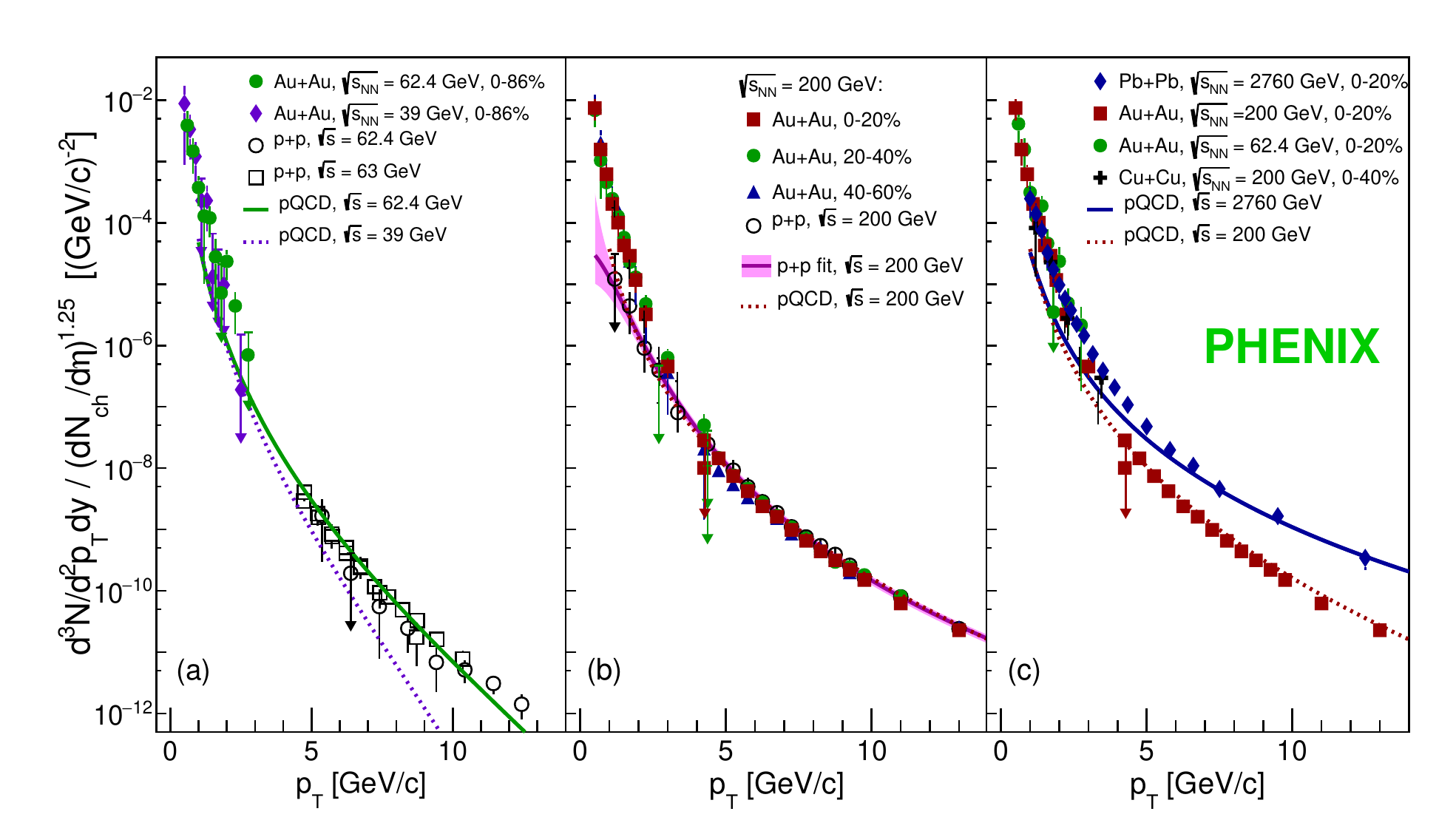}
\caption{Direct photon invariant yields scaled by $(dN_{ch}/d\eta)^{1.25}$
for (a, left) Au+Au collisions at 39 and 62.4\,GeV together with $p$+$p$
and pQCD calculation, (b, middle) Au+Au collisions at 200\,GeV for
several centralities, and (c, right) central Au+Au, Cu+Cu and Pb+Pb
collisions.}
\label{fig5}
\end{figure}

Direct photons are a strong tool to shed a light on the thermodynamics
of the systems, since the photons leave the system unscathed strongly
once emitted. They are also useful for exploring the threshold of partonic matter production. PHENIX has
studied low $p_{T}$ direct photon production for various energies and
collision systems, and found intriguing $dN_{ch}/d\eta$ scaling.
Figure~\ref{fig5} shows the direct photon spectra from large
collision systems scaled by $(dN_{ch}/d\eta)^{1.25}$~\cite{ref7,ref8}.
It shows that the scaled direct photon yield are lying
on top of each other for $p_{T}<5$\,GeV/$c$ that are primarily
soft photons emitted from the bulk system, irrespective of the
collision systems, energies, or centralities.

\begin{figure}[htbp]
\centering
\begin{minipage}{0.41\textwidth}
\centering
\includegraphics[width=0.8\textwidth]{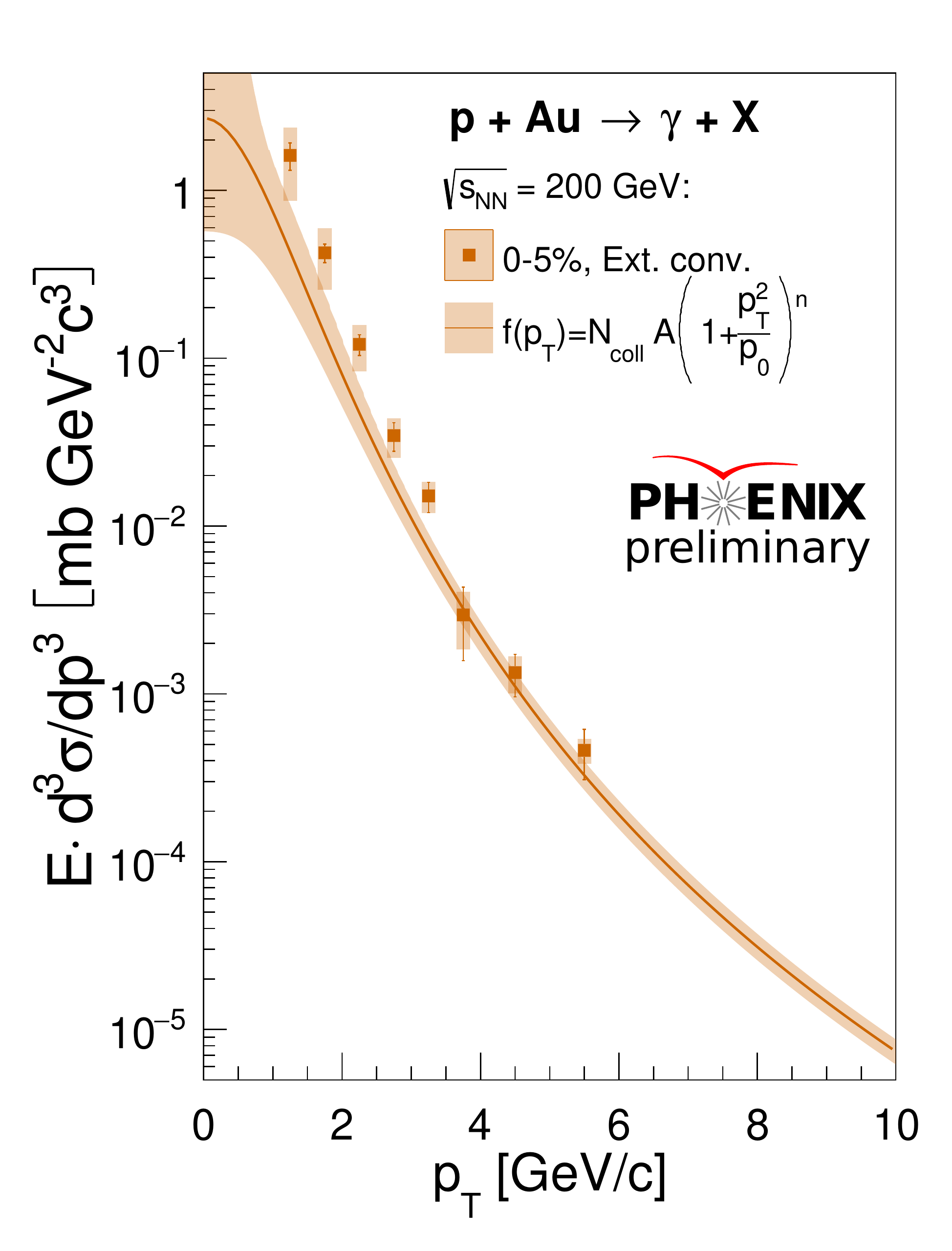}
\end{minipage}
\begin{minipage}{0.55\textwidth}
\centering
\includegraphics[width=0.8\textwidth]{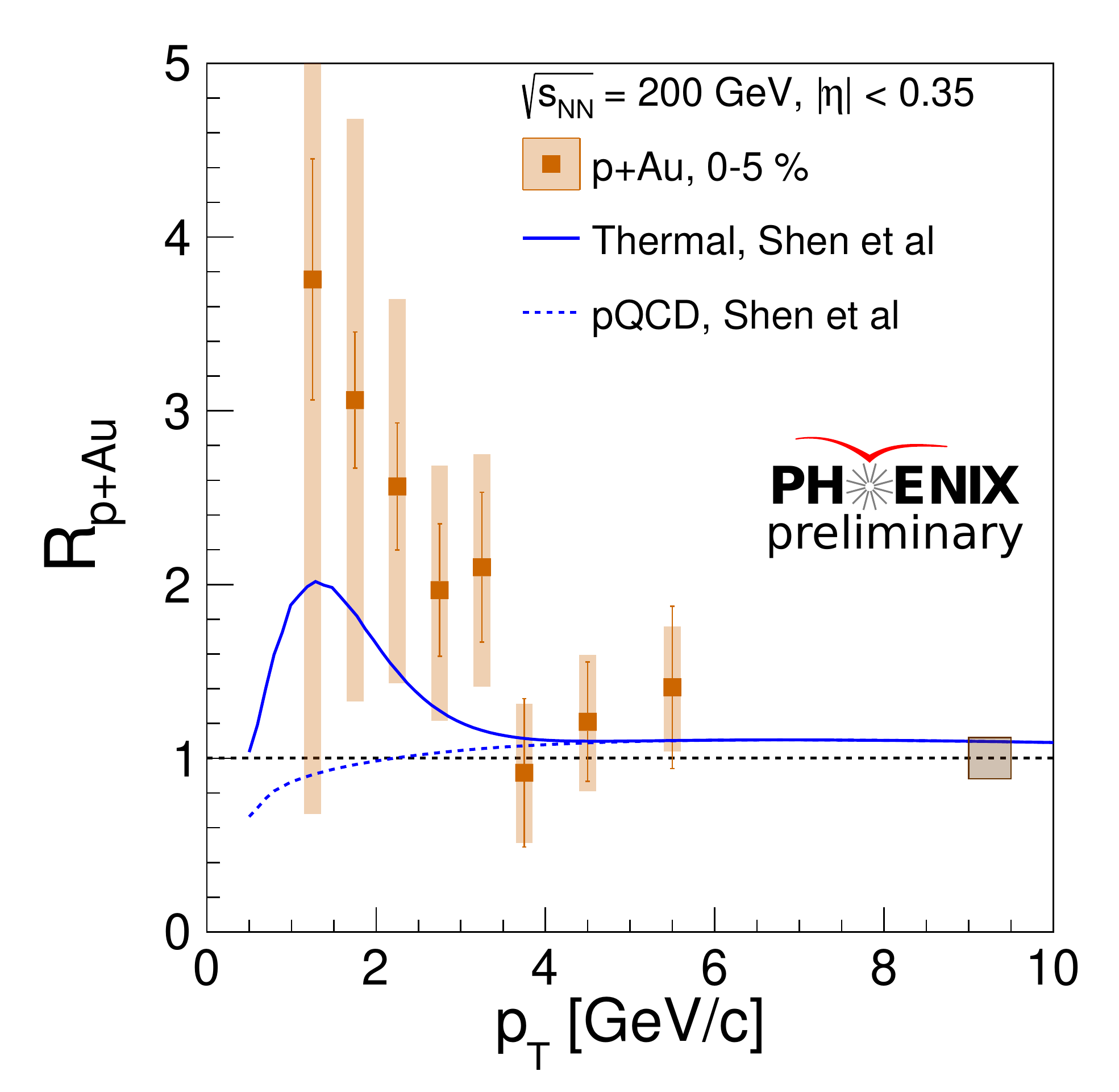}
\end{minipage}
\caption{(a, left) Direct photon spectra in 0-5\,\% $p$+Au collisions
at $\sqrt{s_{NN}}$~=~200\,GeV together with the parameterized $p$+$p$
yield scaled by $N_{\rm coll}$. (b, right) $R_{\rm pA}$ of
the direct photons.}
\label{fig6}
\end{figure}

PHENIX has also measured the direct photons in $p$+Au collisions
at $\sqrt{s_{NN}}$~=~200\,GeV as shown in Fig.~\ref{fig6}.
Although the errors are large, a hint of enhancement over the
expectation from $p$+$p$ collisions is seen. The result is
found to be consistent with
hydrodynamic calculation within errors~\cite{ref9}.

\begin{figure}[htbp]
\centering
%\vspace{-2mm}
\includegraphics[width=0.6\textwidth]{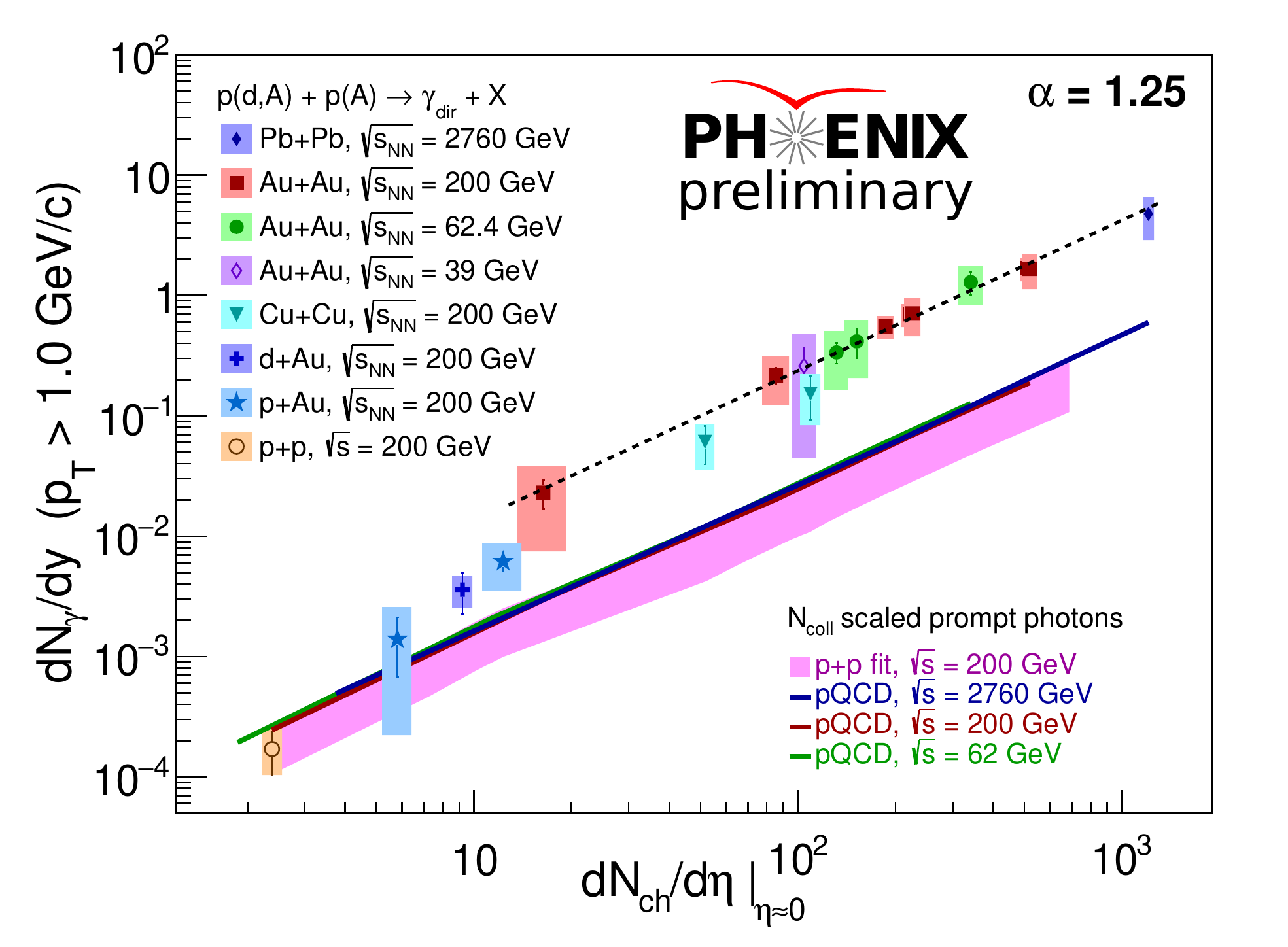}
\caption{Integrated photon yield ($p_{T}>$1\,GeV/$c$) as
a function of $dN_{ch}/d\eta$ for various collision systems.}
\label{fig7}
\end{figure}

We have summarized the direct photon measurements from large to
small systems in the form
of integrated yield ($p_{T}>$1\,GeV/$c$) as a function of
$dN_{ch}/d\eta$ as shown in Fig.~\ref{fig7}.
The dotted line shows the fit to the A+A data with a function of
$dN_{\gamma}/d\eta=\beta (dN_{ch}/d\eta)^\alpha$, where $\alpha$ is
fixed to 1.25. It is found that all the A+A points are on the
dotted lines, while $p$+$p$ and $N_{\rm coll}$ scaled pQCD
calculations are on a different line which is parallel to the
dotted line. The $p/d$+Au data points seem to fill the gap
smoothly between A+A and $p$+$p$ points, which suggests that the
QGP-nization happens smoothly in that $dN_{ch}/d\eta$ range ~\cite{ref8,ref9}. 

\section{Results from small systems}
Since the discovery of collective flow of particles in central
$p$+A collisions at RHIC and the LHC as well as $d/^3$He+Au collisions at RHIC, a question from the
hard probe point of view has been whether or not the nuclear
parton distribution function (nPDF) is strongly modified in
these systems.
\begin{figure}[htbp]
\centering
\begin{minipage}{0.63\textwidth}
\includegraphics[width=1.0\textwidth]{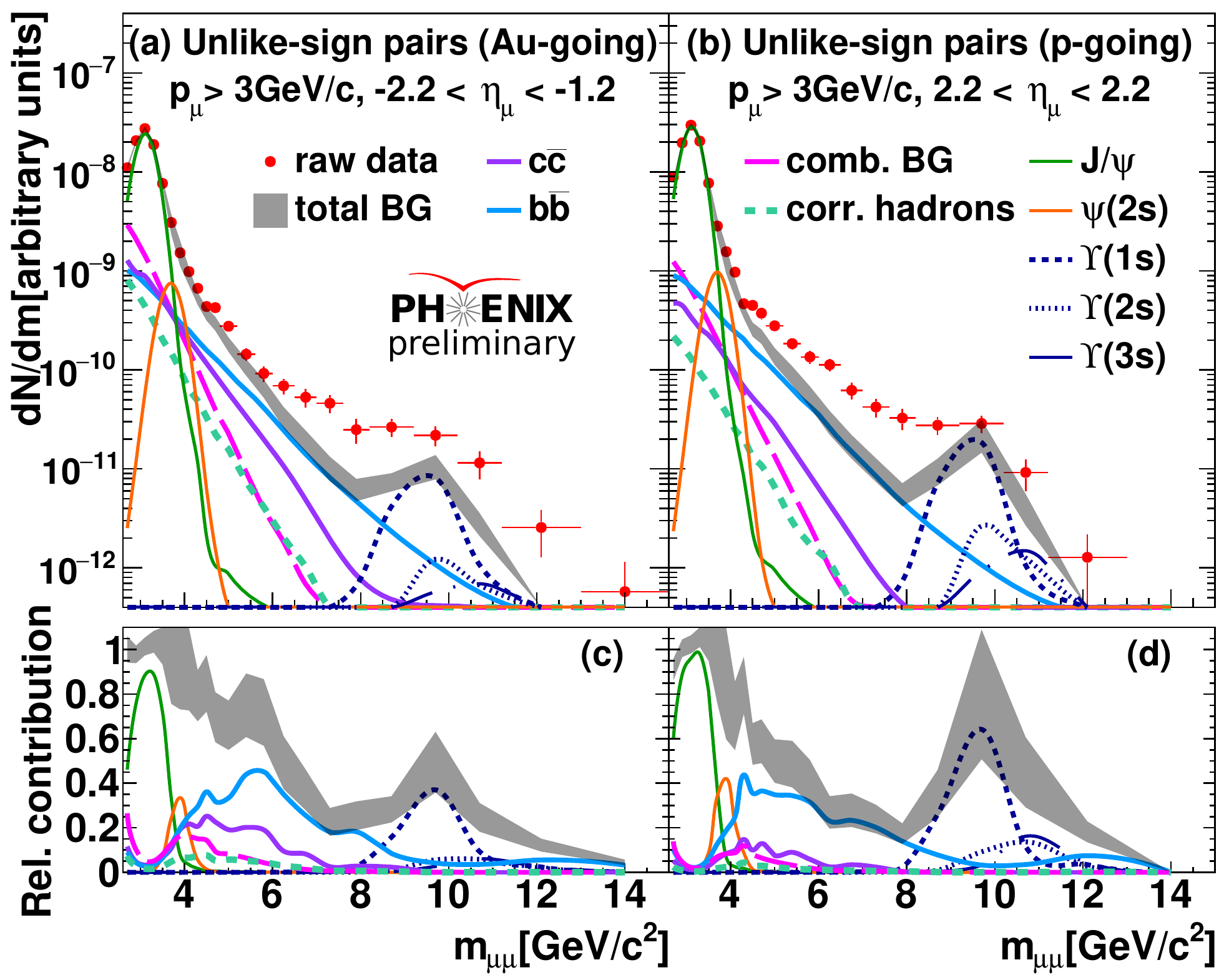}
\end{minipage}
\begin{minipage}{0.34\textwidth}
\includegraphics[width=1.0\textwidth]{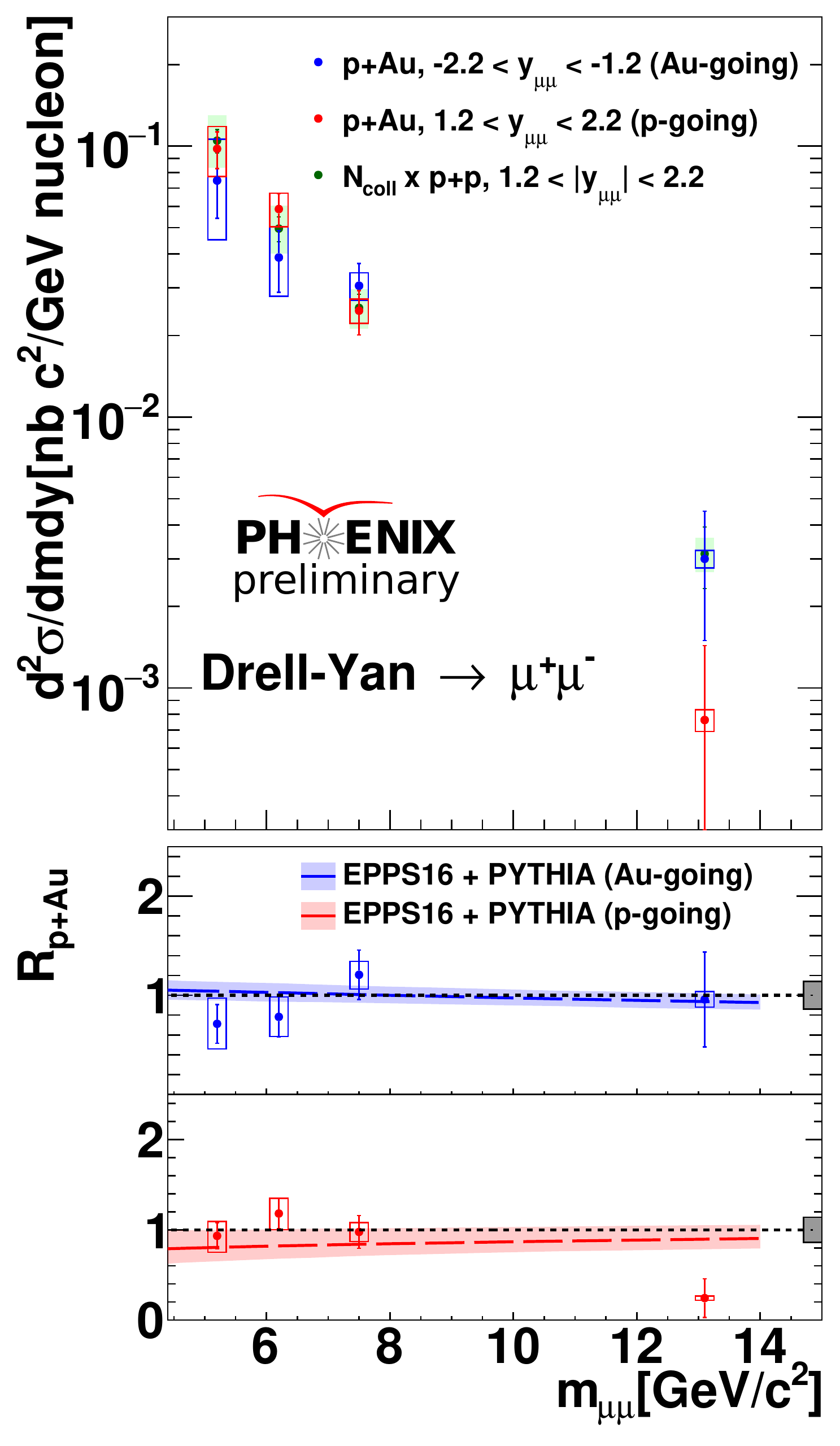}
\end{minipage}
\caption{(a, left) Invariant mass spectra for muon-pairs in
forward and backward rapidities in $p$+Au collisions at
$\sqrt{s_{NN}}$~=~200\,GeV, together with the various known
hadron contributions. (b, right) $p_{T}$ spectra for
the extracted Drell-Yan contribution for $p$+$p$ (scaled by
$N_{\rm coll}$) and $p$+Au collisions, and corresponding $R_{\rm pA}$.}
\label{fig8}
\end{figure}
PHENIX has measured muon-pairs at forward
($p$-going) and backward (Au-going) rapidities
in $p$+Au collisions at
$\sqrt{s_{NN}}$=200\,GeV, and extracted the invariant mass
and $p_{T}$ spectra for the Drell-Yan process, by subtracting
the known hadron decay contribution, as shown in Fig.~\ref{fig8}.
The Drell-Yan process primarily probes the nPDF of the
light quark sector. The $R_{\rm pA}$ shows that the data is
well described by the PYTHIA event generator with the EPPS16
nPDF~\cite{ref10}. With the same dataset but a different
kinematic
cut, we have measured the bottom-quark pair cross-section in
$p$+Au collisions as shown in Fig.~\ref{fig9}.
\begin{figure}[htbp]
\centering
\includegraphics[width=0.75\textwidth]{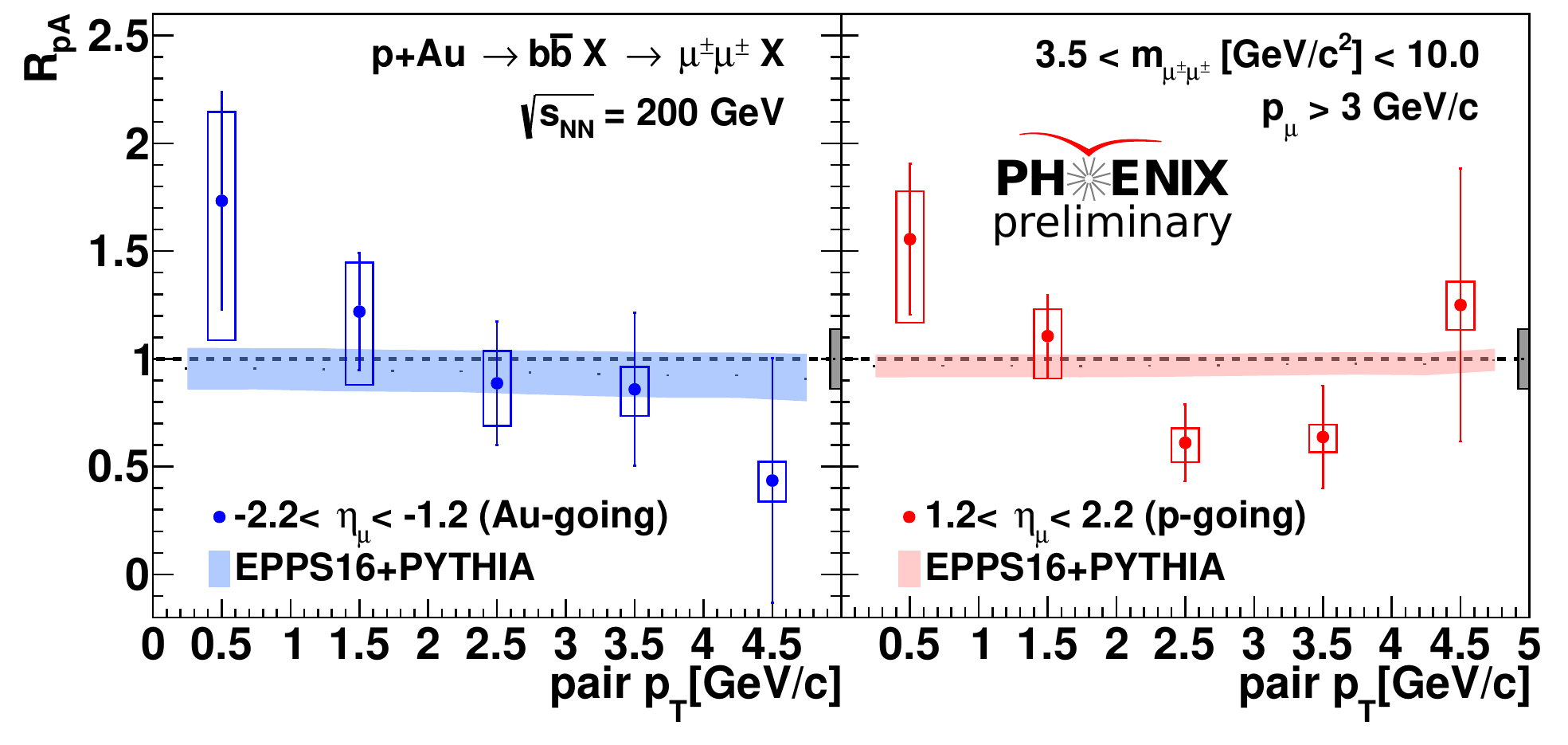}
\caption{$R_{\rm pA}$ of the bottom-quark pairs at forward and
backward rapidities.}
\label{fig9}
\end{figure}
Although the errors are large, the agreement between data and
PYTHIA+EPPS16 nPDF seems to be a bit worse, suggesting that
the gluon part of nPDF has a room to improve since the
bottom-quark pairs are primarily produced from gluons~\cite{ref10}.

\begin{figure}[htbp]
\centering
\begin{minipage}{0.48\textwidth}
\includegraphics[width=1.0\textwidth]{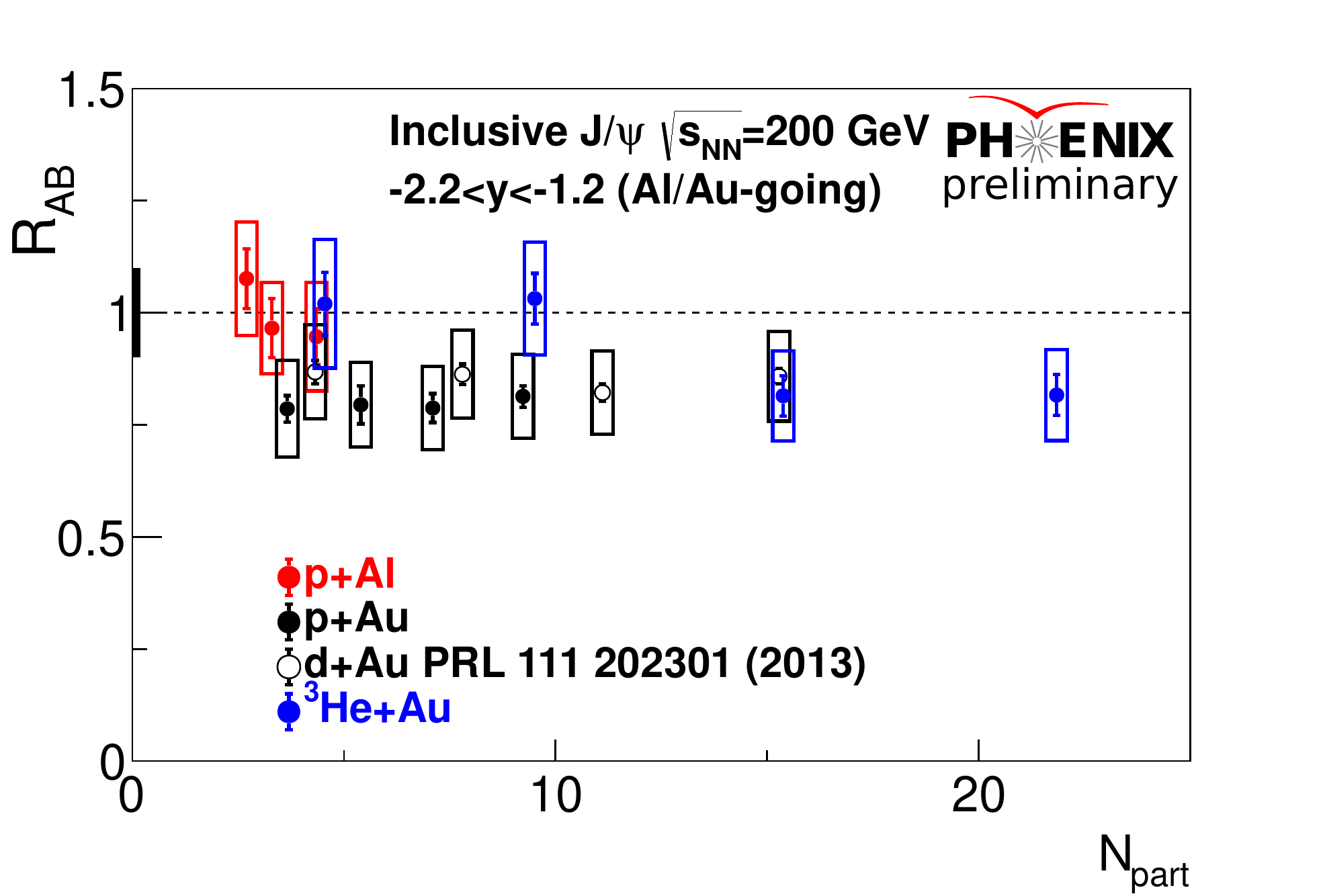}
\end{minipage}
\begin{minipage}{0.48\textwidth}
\includegraphics[width=1.0\textwidth]{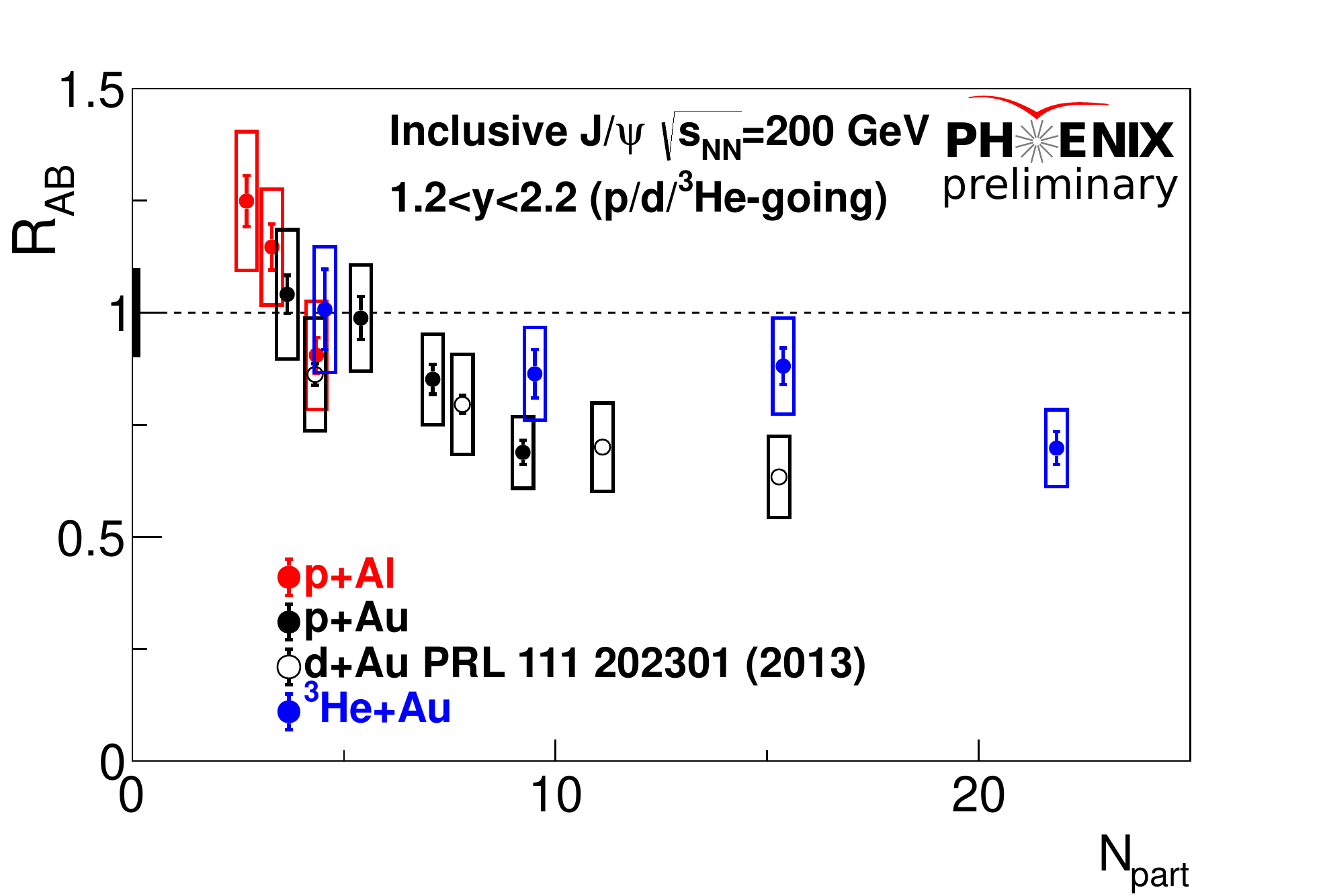}
\end{minipage}
\caption{Inclusive $J/\psi$ $R_{\rm AB}$ at forward at
backward rapidities as a function of $N_{\rm part}$.}
\label{fig10}
\end{figure}

\begin{figure}[htbp]
\centering
\begin{minipage}{0.48\textwidth}
\includegraphics[width=1.0\textwidth]{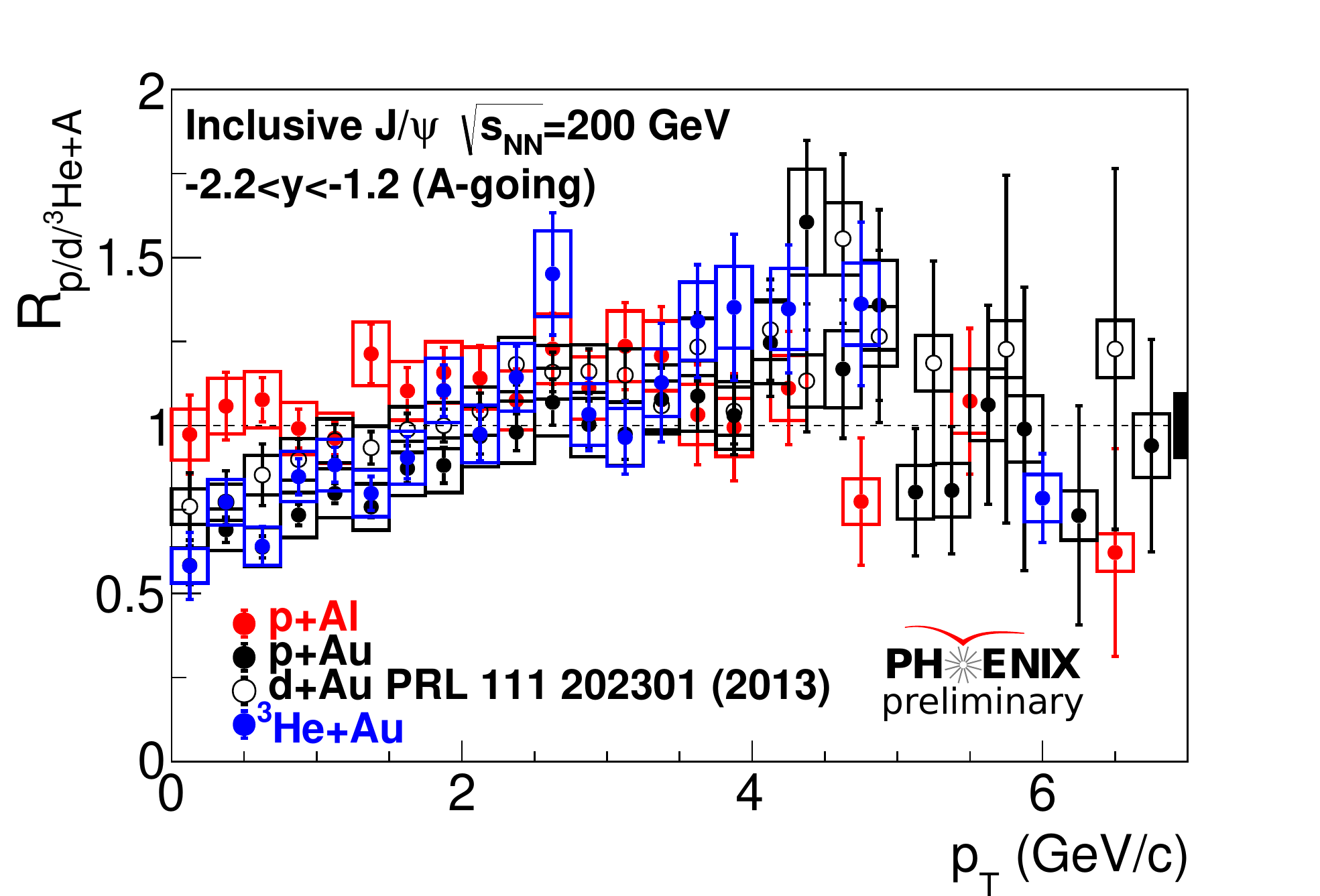}
\end{minipage}
\begin{minipage}{0.48\textwidth}
\includegraphics[width=1.0\textwidth]{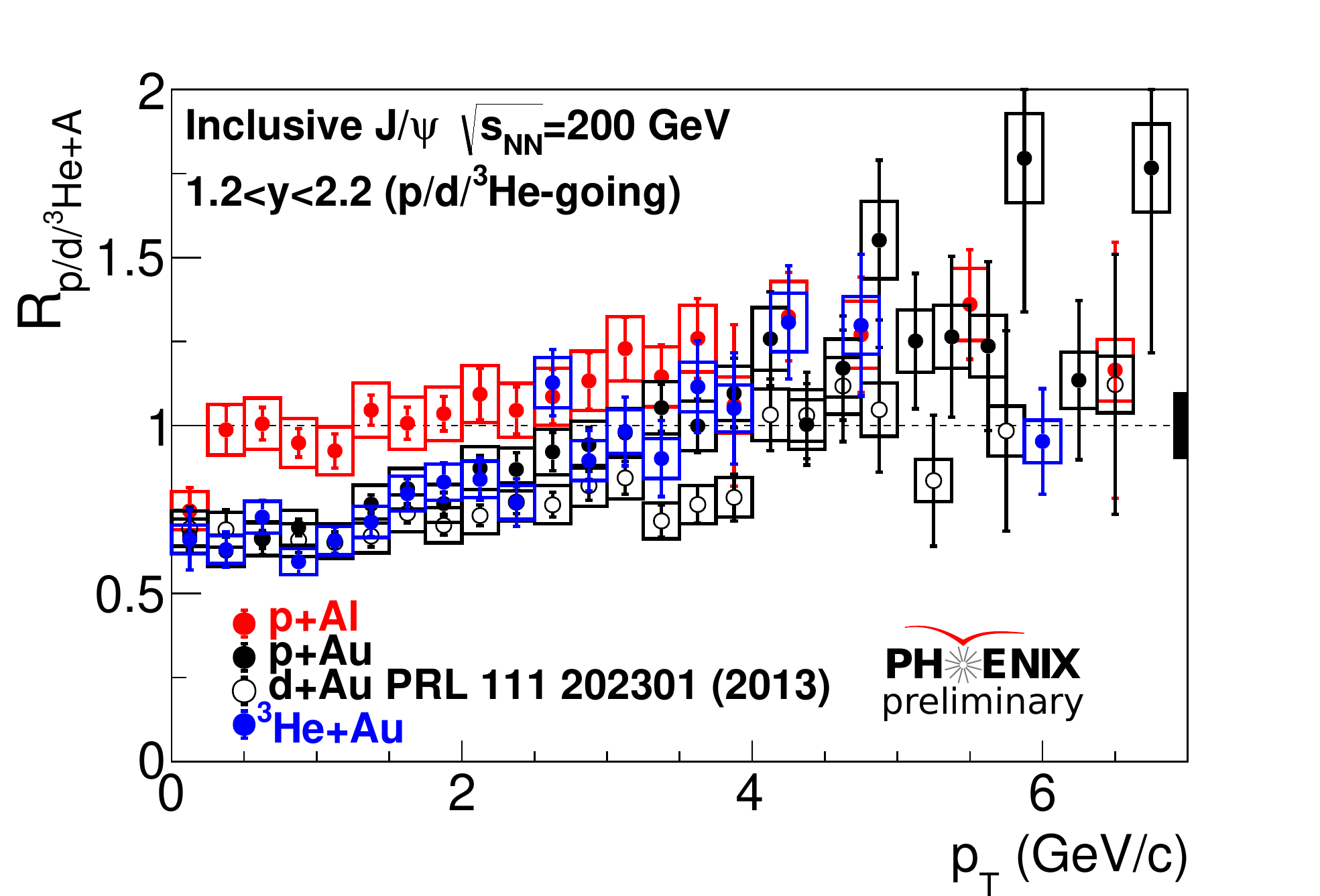}
\end{minipage}
\caption{Inclusive $J/\psi$ $R_{\rm AB}$ at forward at
backward rapidities as a function of $p_{T}$.}
\label{fig11}
\end{figure}

Charmonia provide another handle on nPDF.
%, but will be
%affected by cold nuclear effects at the same time.
PHENIX
has studied $J/\psi$ production in $p$+Al and $p/d/^3$He+Au
collisions at $\sqrt{s_{NN}}$~=~200\,GeV. Figure~\ref{fig10}
shows the inclusive $J/\psi$ $R_{\rm AB}$ at forward
($p/d/^3$He-going) and backward (Au or Al-going) rapidities
as a function of $N_{\rm part}$.
It is found that the $R_{\rm AB}$ scales very well with
$N_{\rm part}$ individually at forward and backward rapidities.
In order to investigate differentially, we have
performed the measurement of $J/\psi$ $R_{\rm AB}$ as a
function of $p_{T}$ as shown in Fig.~\ref{fig11}.
The $R_{\rm AB}$ for $p/d/^3$He+Au collisions are very
consistent each other both at forward and backward
rapidities, while that for $p$+Al collisions
is out of trend, implying the $R_{\rm AB}$ is primarily
determined
by the nPDF or cold nuclear effects of the nucleus~\cite{ref11}.
One thing
worth noting is that the previous single muon measurement
from heavy quarks shows a different trend at backward 
rapidities~\cite{ref12}; single muon $R_{\rm dA}$ is
enhanced, while $J/\psi$ $R_{\rm AB}$
is suppressed. This is consistent with the breakup of $J/\psi$
in the Au nucleus by the cold nuclear effects.

\begin{figure}[htbp]
\centering
\includegraphics[width=0.7\textwidth]{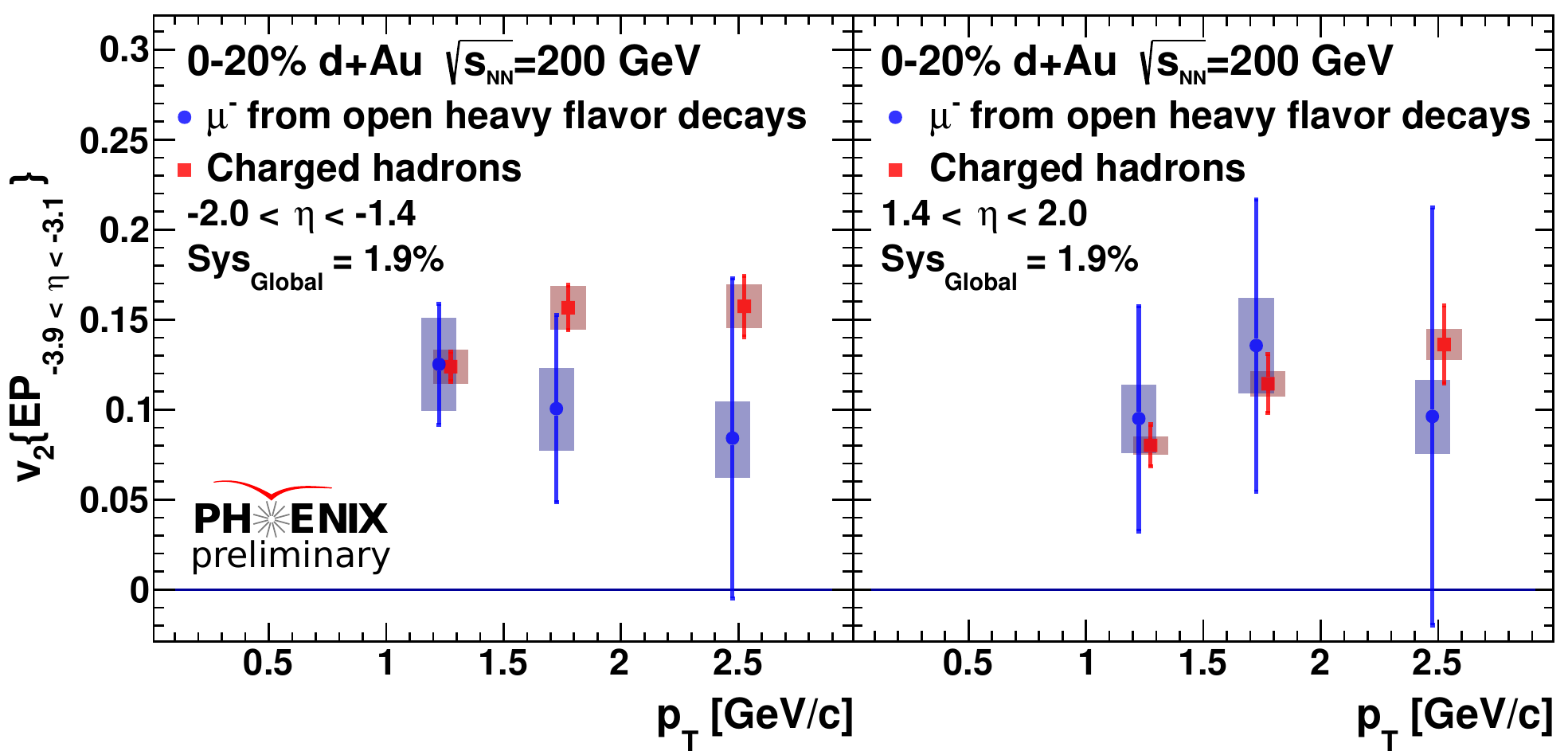}
\caption{$v_2$ of heavy flavor muons and charged hadrons at forward and backward
rapidities in 0-20\,\% $d$+Au collisions.}
\label{fig13}
\end{figure}

Lastly, PHENIX has recently published the collision energy and system
size dependence of light hadron flow ($v_2$ and $v_3$) in the small
collision systems, and found that the results are well described by
hydrodynamic calculations~\cite{ref13}.
PHENIX has also measured the $v_2$ of muons from heavy quarks at
forward ($d$-going) and backward (Au-going) rapidities in most
central $d$+Au collisions at $\sqrt{s_{NN}}$~=~200\,GeV as shown
in Fig.~\ref{fig13}~\cite{ref6}.
Although the errors are large, it was found the muons also flow
and the magnitudes of the flow are consistent with those of
charged hadrons.
%With these results, the production of "mini-QGP" in the small
%systems is rather suggestive, and may be coexistent with the
%conventional cold nuclear effects.
Taken together, these results are rather suggestive of QGP-droplet formation, which does not preclude the coexistence of conventional cold nuclear matter effects in these small systems.

\section{Summary}
We have presented the latest results on the hard probes from large
to small collision systems by fully exploiting the flexibility
of RHIC. The high $p_{T}$ hadrons in large systems are
equally suppressed at the same $N_{\rm part}$, except for
a strange hadron, $\phi$. Away-side jet widths in Au+Au
collisions are found to be larger than and consistent with
those for $p$+$p$ collisions at low and high
associated $p_{T}$, respectively.
The electrons from charm quarks are found to flow more than those
from bottom quarks. This is consistent with more energy loss for
charm quarks.
Soft photon yields ($p_{T}<5$\,GeV/$c$) measured over various
collision systems showed that the
yields scale as: $dN_{\gamma}/dy =\alpha(dN_{ch}/d\eta)^{1.25}$.
The photon yields in $p$+Au collisions look to fill the gap
between the yields in $p$+$p$ and A+A systems when plotting
against $dN_{ch}/d\eta$, hinting a transition from normal to
partonic matter in this $dN_{ch}/d\eta$ region.
Cold nuclear effects probed by Drell-Yan and bottom quarks
are reasonably described by EPPS16 nPDF and PYTHIA event 
generator, modulo a bit worse description for bottom quarks,
suggesting a room of improvement for gluon nPDF.
Comparison of single muon and $J/\psi$ yields in small systems
provided a strong proof of the breakup of $J/\psi$ in the Au
nucleus. Both light-flavor hadrons and heavy quark muons are found to
flow in central $d$+Au collisions, implying that the "mini-QGP" production
in small systems is rather suggestive.

\end{document}